\def\spose#1{\hbox to 0pt{#1\hss}}
\def\lta{\mathrel{\spose{\lower 3pt\hbox{$\sim$}}
    \raise 2.0pt\hbox{$<$}}}
\def\gta{\mathrel{\spose{\lower 3pt\hbox{$\sim$}}
    \raise 2.0pt\hbox{$>$}}}

\documentstyle[12pt,aasms4]{article}

\begin{document}

\title{On The Origin of Cusps in Stellar Systems}
\author{N.W. Evans}
\affil{Theoretical Physics, Department of Physics, 1 Keble Rd, Oxford,
OX1 3NP, UK}
\and
\author{J.L. Collett}
\affil{Division of Physics and Astronomy, University of Hertfordshire,
Hatfield, AL10 9AB, UK}

\begin{abstract} An origin is sought for the ubiquity of 
cusps, both in computer simulations of halo formation in hierarchical
clustering cosmogonies and in observations of galactic nuclei by the
Hubble Space Telescope (HST). The encounters of merging clumps that
built the galaxies can be described by the collisional Boltzmann
equation. Using insights gained by studying the simpler Fokker-Planck
equation, we show that there is a steady-state, self-consistent,
cusped solution of the collisional Boltzmann equation corresponding to
$\rho \sim r^{-4/3}$. This equilibrium is both stable and an
attractor. It is the natural end-point of the diffusive encounters of
an ensemble of equal mass clumps.  The introduction of a mass spectrum
weakens the mass density cusp. The spike in the luminosity density can
be accentuated or softened, depending on the form of the
mass-luminosity relation. Possible applications to the cusped nuclei
of early-type galaxies are discussed.

\end{abstract}


\section{INTRODUCTION}

Over the last years, two strands of research have made galactic
astronomers more aware of the importance of cusps.  First, Navarro,
Frenk \& White (1996a) have presented evidence for the existence of a
universal law for galaxy halos, in which the mass density $\rho$ is
cusped and behaves like $r{}^{-1}$ at small radii. This provides a
reasonably accurate description of the density law for all halo masses
in a range of cosmologies with different global parameters and power
spectra (Navarro, Frenk \& White 1996b). Fukushige \& Makino (1996)
have run simulations with larger numbers of particles and smaller
softening. They report that the cusp is steeper than $r{}^{-1}$ and
that the central regions of the halo are expanding. Secondly, there is
an extensive data-set of images of galactic nuclei taken with the HST
by Lauer et al. (1995). The brilliance of the central nuclei is a
consequence of a power-law rise in the light distribution right down
to the very limits of the resolution.  The data-set has been analysed
by Gebhardt et al. (1996), who argue for a bimodal distribution of
cusp slopes of early-type galaxies.  Giant, radio-loud, early-type
galaxies have a luminosity density $\nu$ that is typically cusped like
$r{}^{-0.8}$, whereas normal or dwarf, radio-quiet, early-type
galaxies have much steeper power-law profiles like $r^{-1.9}$. Hardly
any early-type galaxies are uncusped.  Of course, there is more than
one mechanism for the formation of stellar cusps (e.g., Bahcall \&
Wolf 1976; Faber et al. 1996; Syer \& White 1996). The aim of this
paper is to point out another possibility and discuss its likely
applications.

\section{CUSPED STEADY-STATE SOLUTIONS OF THE FOKKER-PLANCK AND
COLLISIONAL BOLTZMANN EQUATIONS}

Ever since the visionary contribution of Toomre \& Toomre (1972), the
importance of merging and accretion in the assemblage of elliptical
galaxies and the halos of spiral galaxies has been widely
acknowledged. The building of the galaxies involves both violent
relaxation and the encounters or collisions of merging clumps. A
complete description of such collisional dynamics is offerred by
Gilbert (1968), albeit at the price of some mathematical
complexity. Gilbert showed that the collisional effects may be divided
into two distinct processes, systematic acceleration from polarisation
clouds (dynamical friction) and diffusion. It therefore seems a
reasonable first approach to study the dynamics with the simpler
Fokker-Planck equation, which includes the essential physical
processes. This equation models the energy changes of the clumps as
the cumulative effects of many weak binary encounters. This is
probably invalid in the cosmogonic applications we have in
mind. Nonetheless, we shall see that certain steady-states of the
Fokker-Planck equation hold good for the collisional Boltzmann
equation, which fully incorporates the effects of strong encounters.

For spherically symmetric, isotropic galaxies, the Fokker-Planck
equation reads (e.g., Spitzer 1987; Theuns 1996)
\begin{equation}
\label{eq:fp} 
\Bigl( {\partial N(E,t) \over \partial t} \Bigr)_{\rm enc} =
-{\partial \over \partial E}[ N(E,t)D_1(E)] + {1\over 2}{\partial^2
\over \partial E^2}[N(E,t)D_2(E)]
\end{equation}
Here, $N(E,t) dE$ is the number of clumps with energy in the range $E$
to $E + dE$ at time $t$. The first and second terms on the right-hand
side describe the effects of dynamical friction and diffusion
respectively. For a clump of mass $M$ moving through a system of equal
mass clumps, the diffusion coefficients $D_1$ and $D_2$ are defined as
(e.g., Spitzer 1987; Theuns 1996)
\begin{equation}
\label{eq:done} 
D_1(E) = \langle \Delta E \rangle_{\rm V} = 16 \pi^2 G^2 M^2 \log \Lambda 
\Bigl[ \int_E^\infty f(E') dE' - {1\over p(E)}\int_0^E f(E')p(E') dE'
\Bigr],
\end{equation} 
\begin{equation}
\label{eq:dtwo} 
D_2(E) = \langle (\Delta E )^2\rangle_{\rm V} = 32 \pi^2 G^2 M^2 \log \Lambda 
\Bigl[ {1\over p(E)}\int_E^\infty q(E') f(E') dE' + {q(E)\over p(E)}
\int_0^E f(E') dE' \Bigr].
\end{equation}
Here, angled brackets denote averages over all clumps of energy $E$
within the accessible volume V. The $\log \Lambda$ term is the
\lq Coulomb logarithm', while $f$ is the phase space number density
of the clumps.  The quantity $p(E)$ is the density of states (Spitzer
1987, chap. 2)
\begin{equation}
p(E) = 16 \pi^2 \int_0^{r_{\rm max}(E)} [2(E - \phi (r))]^{1/2} r^2 dr,
\end{equation}
while $q(E)$ is the total phase space volume with energy less than $E$ 
\begin{equation}
q(E) = {16 \pi^2 \over 3} \int_0^{r_{\rm max}(E)} 
[2(E - \phi (r))]^{3/2} r^2 dr.
\end{equation}
In both these definitions, $r_{\rm max} (E)$ is the maximum radial
excursion possible for a clump of energy $E$, while $\phi$ is the
gravitational potential.

Are there any cusped, steady-state solutions of the Fokker-Planck
equation for our crude facsimile of a proto-galaxy built from merging
clumps? Suppose the potential and density have the scale-free form
\begin{equation}
\label{eq:scalefree}
\phi \sim r^{-\beta},\qquad \rho \sim r^{-(\beta+2)},\qquad \beta \neq 0,
\end{equation}
where $\beta$ is a constant describing the severity of the
cusp. The isotropic phase space distribution $f(E)$ and the density
of states $p(E)$ behave like (e.g., Evans 1994)
\begin{equation}
f(E) = f_0 E^{2/\beta - 1/2},\qquad p(E) = p_0 E^{1/2 - 3/\beta},
\end{equation}
where $f_0$ and $p_0$ are constants. So, the number of clumps $N(E)$
with energy in the range $E$ to $E + dE$ goes like
\begin{equation}
N(E) = f(E) p(E) =f_0 p_0E^{-1/\beta}.
\end{equation}
It is now straightforward to deduce the diffusion coefficients $D_1$
and $D_2$ as
\begin{equation}
D_1(E) = 16\pi^2 G^2 M^2 f_0 \log \Lambda \,  {\beta (2+3\beta)\over
(\beta+4)(1-\beta)} \, E^{1/2 + 2/\beta},
\end{equation}
\begin{equation}
D_2(E) = 64 \pi^2 G^2 M^2 f_0 \log \Lambda \, {\beta^2\over (\beta +4)
(1-2\beta)}\, E^{3/2 + 2/\beta}.
\end{equation}
Substituting into the Fokker-Planck equation (\ref{eq:fp}), we find
that a stellar cusp can be in a steady-state under the effects of
encounters if
\begin{equation}
\Bigl( {\partial N(E,t) \over \partial t} \Bigr)_{\rm enc} =
8\pi^2 G^2M^2 f_0^2 p_0 \log \Lambda 
{\beta (3\beta +2)(\beta +2)\over (\beta +4)(1-\beta)(1
- 2\beta)} E^{1/\beta - 1/2} = 0.
\end{equation}
Note that the case $\beta =0$ corresponds to an isothermal cusp with $\rho
\sim r^{-2}$. This is a steady-state solution, but it requires a special 
treatment as the gravitational potential is logarithmic rather than a
power of radius.  The other physical solution corresponds to $\beta =
-2/3$ or
\begin{equation}
\label{eq:magic}
\rho \sim r^{-4/3}.
\end{equation}
This satisfies the steady-state Fokker-Planck equation in an
interesting way. The dynamical friction and diffusive flux terms on
the right-hand side of (\ref{eq:fp}) separately vanish.

Now, the Fokker-Planck equation is derived from the full collisional
Boltzmann equation (e.g., Spitzer 1987)
\begin{equation}
\Bigl( {\partial N(E,t) \over \partial t} \Bigr)_{\rm enc} =
\sum_{r =1}^\infty {(-1)^r \over \Gamma(r+1)}{\partial^r \over \partial E^r}
[ N\langle (\Delta E )^r \rangle_{\rm V} ].
\end{equation}
under the assumption that the velocity changes produced by the
encounters are small and so only the first two terms in the Taylor
expansion matter. This is unlikely to be the case in our cosmogonic
application. But, directly from scaling arguments, the diffusion
coefficients must behave like
\begin{equation}
D_r = \langle (\Delta E )^r \rangle_{\rm V} \sim E^{2/\beta + r - 1/2}.
\end{equation}
When $\beta = -2/3$, it follows that $N \langle (\Delta E )^r 
\rangle_{\rm V} \sim E^{r-2}$. This means that every term in the Taylor
expansion -- bar the first -- is annihilated. But, the first term
vanishes because $D_1 =0$. So, we are led to an astonishing
conclusion: the stellar cusp with $\rho \sim r^{-4/3}$ is a
steady-state solution not just of the Fokker-Planck equation, but also
of the full collisional Boltzmann equation. This vindicates our
earlier claim that an analysis of the Fokker-Planck equation may be a
good guide to the collisional Boltzmann equation. The time-scale on
which the effects of Fokker-Planck evolution become observable is the
two-body relaxation time, which can be longer than the age of the
Universe for a giant galaxy. Again, though, let us stress that we are
merely using the Fokker-Planck equation as a guide to the
harder-to-handle collisional Boltzmann equation. Large-angle
scattering of clumps by potential fluctuations can drive evolution on
a much shorter time-scale.

This simple treatment may be refined by introducing a distribution
of masses of the clumps. Suppose the phase space number density 
of lumps with mass in the range $M$ to $M + dM$ is (c.f. Young (1977))
\begin{equation}
\label{eq:tobeint}
f(E,M) \sim M^{X-1} \exp[ -ME^{\alpha/(2 + X)} ],
\end{equation}
where $\alpha$ and $X$ are constants. This form of the distribution
function is attractive because, once the integration over all the
masses has been performed, the phase space mass density is again a
power of the energy.  As $X \rightarrow \infty$, the distribution
function reduces to a simple power of the energy for a single-mass
species.  If the potential and density are of the scale-free power-law
form (\ref{eq:scalefree}), then integrating (\ref{eq:tobeint}) over
all masses and energies to obtain the mass density in configuration
space implies
\begin{equation}
\alpha = {2 + X \over 1 + X}\,{\beta -4 \over 2\beta}.
\end{equation}
It is a straightforward, but lengthy, calculation to work out the new
diffusion coefficients. We find that -- for any value of the mass
spectrum parameter $X$ -- there is a cusped, steady-state solution of
both the Fokker-Planck and the collisional Boltzmann equations with
\begin{equation}
\rho \sim r^{-{\displaystyle 4\over \displaystyle 3} + {\displaystyle 14\over
\displaystyle 3(2+3X)}}.
\end{equation}
In the single species limit, $X \rightarrow \infty$ and we recover
$\rho \sim r^{-4/3}$. The effect of broadening the mass distribution
is to weaken the severity of the cusp.  To compare our solutions with
the HST data on early-type galaxies, the important quantity is not the
mass density, but the luminosity density.  Suppose the mass-luminosity
relation behaves like $L(M) \propto M^p$, then the luminosity density
is cusped like
\begin{equation}
\nu \sim r^{-{\displaystyle 4\over \displaystyle 3} - {\displaystyle
7(3p-5)\over \displaystyle 3(2+3X)}}.
\end{equation}
If $p > 5/3$, then the effect of the distribution of masses is to make
the luminosity density cusp steeper than $r^{-4/3}$.  Notice, too,
that when most of the mass in the cusp is composed of dark, heavy
remnants, then the luminosity spike is less steep than $r^{-4/3}$.  If
there is a distribution of masses, then the heavier objects settle
deeper in the cusp -- and so the luminosity spike can be enhanced or
diminished, according to whether the heavier objects are luminous or
dark.

What are the merging clumps? One possibility is that the clumps are
individual stars. This is fine, as the diffusion coefficients
(\ref{eq:done}) and (\ref{eq:dtwo}) are derived for an ensemble of
point masses. Another possibility is that the clumps are bound
clusters of stars. Now, though, the diffusion coefficients should
strictly speaking be modified to take account of the extended
structure of the clumps. A third possibility is to identify the clumps
with gas clouds that have not yet formed stars. This also requires
modification of the diffusion coefficients to allow for the
dissipative nature of lossy encounters between gas clumps.

\section{CUSP THERMODYNAMICS AND STABILITY}

Our solutions are in dynamic -- but not thermodynamic --
equilibrium. This is obvious as the only possible thermodynamic
equilibrium for a stellar system is the isothermal sphere. So, even
though the mass flux vanishes in our Fokker-Planck solutions, there is
a heat flux transmitted from the exterior of the model to the center
of the cusp.  Defining the thermodynamic temperature $T(E,M)$ as
(e.g., Inagaki \& Lynden-Bell 1990)
\begin{equation}
[kT(E,M)]^{-1} = -{1\over M}{\partial \log f (E,M)\over \partial E}
\end{equation}
then a nice picture to have in mind is the conduction of heat from
the hotter outsides to the colder center of the cusp.  In general, the
heat flux $S(E)$ through phase space volume $q(E)$ is defined by 
(H\'enon 1961; Lynden-Bell 1996)
\begin{equation}
{1\over 16 \pi^2 G^2M^3 \log \Lambda}{d S \over dE} =
f(E)\int_0^E f(E'){dq\over dE'} dE' + {df\over dE}\Bigl[
\int_0^E f(E')q(E') dE' + q(E)\int_E^\infty f(E')dE'\Bigr].
\end{equation}
For power-law cusps of the form (\ref{eq:scalefree}), we readily find
\begin{equation}
S(E) \sim E^{3/2 + 1/\beta}.
\end{equation}
So, for the single-species Fokker-Planck solution with $\beta = -2/3$,
there is a constant heat flux from the outside to the interior. It has
magnitude
\begin{equation}
S = {25\pi \over 768\sqrt{2}} \log\Lambda {M \phi_a^{3/2} \over a},
\end{equation}
where $\phi_a$ is the gravitational potential at some reference radius
$a$. There is an evident analogy here with the earlier work of Bahcall
\& Wolf (1976), who considered the collisional relaxation of stars in
the potential of a central massive black hole. They too found a steady
state with a constant heat flux maintaining the equilibrium by having
the hole absorb the influx of energy. To sustain the self-consistent
cusp, an analogous sink is required.  What are the physical causes of
this sink of kinetic energy in the center of the self-consistent
Fokker-Planck cusps?  There are a number of possibilities. In
cosmogonic applications, the disruption of binary or multiple lumps
(such as weakly bound clusters) can provide an energy sink. Softly
bound objects absorb energy and eventually disassociate according to
Heggie's (1975) Law. This effect is likely to be more important in
galaxy halo formation than, for example, in the present-day evolution
of the globular clusters. In dense galactic nuclei, collisions should
affect a large fraction of stars (Davies 1996). The disruption of
stars through coalescence or collision again provides a sink of
kinetic energy -- both in the direct impact of stars and via the
raising of tides.  This, though, is a delicate matter. If the density
is high enough for collisions to be important, then hard binaries will
also form. They will act as a heat source, counteracting the effects of
collisions.

Is the self-consistent cusp stable? The evidence for stability is the
series of numerical computations of the evolution of the Fokker-Planck
equation that we have performed and will present elsewhere. The cusp
remains unchanged over time-scales longer than the age of the
Universe. When the Fokker-Planck equation is evolved with arbitrary
initial conditions, the distribution function moves towards the
steady-state cusp, which is therefore an attractor. These properties
are analogous to those found by Bahcall \& Wolf (1976) for cusped star
distributions around a black hole. At first glance, though, they seem
in contradiction with the numerical results of Quinlan (1996), who
found that the double-power-law models evolve quickly (by expanding)
when the inner cusp is $\rho \sim r^{-4/3}$. But, Quinlan's
simulations do not incorporate a central heat sink and so his
interesting results are not directly applicable to our problem.

\section{DISCUSSION}

Navarro, Frenk \& White (1996a,b) have recently argued that dark halos
possess a universal density profile of the form
\begin{equation}
\label{eq:nfw}
\rho \sim {1 \over r(a+r)^2}
\end{equation}
In the inner parts of their simulations, the dynamics is driven by
encounters. There is quite a lot of scatter in the density profile in
the inner region of the simulations, which may still be affected by
numerical limitations. So, the inner cusp profile could be $\rho \sim
r^{-4/3}$ without violating the simulation data (White, private
communication). This idea receives support from the higher-resolution
simulations of Fukushige \& Makino (1996), who find both that the cusp
is steeper than $\rho \sim r^{-1}$ and that the central regions of the
haloes are expanding. This is exactly what we would predict on
theoretical grounds. Heat has to be transferred from the outer to the
inner parts of the N-body simulation, which cannot be in exact
thermodynamic equilibrium. In the absence of an energy sink, the
central regions must respond to the heat input by expansion.  The
intrinsic velocity dispersion of our cusp solutions falls on moving
deeper into the cusp. This property also appears to be replicated in
the simulations (White, private communication; Fukushige \& Makino
1996).

Faber et al. (1996) argue that the shallow cusps of giant ellipticals
may be caused by black hole binaries, but that the steep cusps of
normal or dwarf ellipticals may be the result of the engulfment of
gas-rich, small disk galaxies.  Here, we raise the question: is it
possible that some of the cusps correspond to these new steady-state, 
attracting solutions?  This is difficult to establish
for certain -- as the light cusp depends on both the mass spectrum and
the mass-luminosity relation. These are poorly known for the
stars in the cusps.  If the light in the cusps is dominated by main
sequence stars, then a fair value for $p$ is $\sim 3$ so that $L
\propto M^3$. Then, the observed luminosity density spike of the
radio-quiet early type galaxies can be reproduced if $X \sim 4$, which
corresponds to a broadish mass spectrum. The mass density cusp is just
$\rho \sim r^{-1}$ and the intrinsic velocity dispersion $\langle v^2
\rangle$ vanishes as $r \rightarrow0$. Assuming the density cusp is
matched to the outer parts of a model like (\ref{eq:nfw}), then the
line-of-sight velocity dispersion falls logarithmically to zero at the
centre.  Does the kinematic evidence support this suggestion? The
available ground-based data seem to show the line-of-sight velocity
dispersion is constant or rising in the cusped nuclei of the sample
(Faber et al. 1996). But, the inference of the velocity dispersion is
a delicate matter in the centers of galactic nuclei where the line
profile is non-Gaussian. If mass segregation of a stellar population
operates to build the luminosity density cusp, then color gradients
will be established. The available data is again scanty, but the
general belief is that large color gradients are not present (Carollo
et al. 1997). This, though, is not a serious objection. It is easy to
evade the existence of colour gradients with more complex
mass-luminosity relations than the simple power-laws -- for example,
if the cusp is built from a uniform stellar population and lower-mass
black holes or dark remnants, mass segregation can build the observed
luminosity spike, but no color gradients will be
detected. Alternatively, if the constituents of the cusp are a uniform
stellar population and higher mass black holes, then the spike in the
luminosity density will be less than $\nu \sim r^{-4/3}$ with no color
gradients.

The main result of this paper is the demonstration of a new cusped
family of self-consistent solutions to the Fokker-Planck and
collisional Boltzmann equations. These are a natural end-point of the
gravitational scattering of point masses in the presence of a central
energy sink.  Central light cusps are possible in the basin of a
potential well without central massive objects.

\acknowledgments

NWE is supported by the Royal Society, while JLC acknowledges
financial support from PPARC (grant number GRJ 79454). We wish to
thank James Binney, Marci Carollo, Tim de Zeeuw, Suvendra Dutta,
Man-Hoi Lee, Donald Lynden-Bell, Gerry Quinlan, Massimo Stiavelli, Tom
Theuns, Scott Tremaine, Roeland van der Marel and Simon White for
helpful discussions on these matters. We particularly wish to thank
Donald Lynden-Bell for access to unpublished notes.

\end{document}